\documentclass[10pt,letterpaper]{article}

\usepackage[matrix,arrow,ps]{xy}
\usepackage{opex3}

\usepackage{amsmath,amsfonts}
\usepackage{color}
\usepackage{amssymb}
\usepackage{amsfonts}

\def\N{\mathbb{N}}

\newcommand{\beq}{\begin{eqnarray}}
\newcommand{\eeq}{\end{eqnarray}}

\begin{document}

\title{Revolution analysis of three-dimensional arbitrary cloaks}

\author{Guillaume Dupont,$^{1}$ S\'ebastien Guenneau,$^{1}$ Stefan Enoch,$^{1}$ Guillaume Demesy,$^{1}$
Andr\'e Nicolet$^{1}$, Fr\'ed\'eric Zolla$^{1}$ and Andr\'e
Diatta,$^{2}$}

\address{$^{1}$Institut Fresnel, UMR CNRS 6133, Universit\'e Aix-Marseille III, 13397 Marseille, France}
\address{$^{2}$Department of Mathematical Sciences, Peach Street, Liverpool L69 3BX, UK}
\email{sebastien.guenneau@fresnel.fr}

%
%



\begin{abstract}
We extend the design of radially symmetric three-dimensional
invisibility cloaks through transformation optics \cite{pendrycloak}
to cloaks with a surface of revolution. We derive the expression of
the transformation matrix and show that one of its eigenvalues
vanishes on the inner boundary of the cloaks, while the other two
remain strictly positive and bounded. The validity of our
approach is confirmed by finite edge-elements
computations for a non-convex cloak of varying thickness.
\end{abstract}

\ocis{(000.3860) Mathematical methods in physics; (260.2110)
Electromagnetic theory; (160.3918) Metamaterials; (160.1190)
Anisotropic optical materials}


\section{Introduction}
It was recently found that transformation optics open new avenues in
electromagnetic cloaking, either through their heterogeneous
anisotropic effective material parameters (transformation optics,
\cite{pendrycloak,leonhardt}) or through low index materials
\cite{engheta} or negative refractive index materials \cite{graeme}.
Interestingly, the invisibility is preserved in the case of an
intense near field \cite{opl2007}, when the ray optics picture
breaks down. The mathematics behind the scene have been known from
researchers working in the area of inverse conductivity problems
\cite{greenleaf}. The first experimental realization of an
invisibility cloak, chiefly achieved in the microwave regime
\cite{science}, suggests that cloaking will be limited to a very
narrow range of frequencies. However, it will not be perfect since
the cloak is necessarily dissipative and dispersive, and some of its
tensor components are singular on the inner boundary. The latter
drawback can be overcome by considering a generalized transform
\cite{heapl,tyc} in an upper dimensional space and then projecting
the resulting metric on the physical space, and this leads to
non-singular tensors of permittivity and permeability.
Alternatively, one can design an approximate structured cloak via
homogenization \cite{naturecloak,farhat,fengprl}, although the rapid
growth of the field, fueled by a keen interest of the optics
community, promises a large panel of new technological applications.

A non trivial question to ask is whether one can design cloaks of
non-spherical shapes. A parameterization of the cloak's boundaries
was proposed by three of us to design cylindrical cloaks of an arbitrary cross-section
\cite{opl2008}.
Its extension to the general three-dimensional case now requires to
parameterize the inner and outer boundaries of the cloak as some surfaces with varying
radii

\begin{equation}
\rho(\theta,\phi) = a_{0,0} + \displaystyle
\sum_{(m,n)\in\N^2\setminus\{(0,0)\}} \{ a_{m,n} \cos(m\theta+n\phi)+
b_{m,n}\sin(m\theta+n\phi) \} \; ,
\end{equation}
which is somewhat out of reach running the COMSOL multiphysics
package on a $256$ Gb RAM computer: Compared to a spherical cloak,
the tetrahedral mesh needs be further refined due to the complexity
of surfaces involved. However, to demonstrate the versatility of our
Fourier-based approach, it is enough to analyse arbitrary cloaks
built with surfaces of revolution i.e. generated by rotating a curve
about an axis.

In the present paper, we discuss the design of axially invariant three-dimensional
cloaks with an arbitrary cross-section described by two functions
$R_1(\phi)$ and $R_2(\phi)$ giving an angle dependent distance from
the origin.  These functions correspond respectively to the interior
and exterior boundary of the cloak. We shall only assume that these
two boundaries can be represented by a differentiable function.
Their finite Fourier expansions are thought in the form $R_j(\phi) =
a^j_{0,0} + \sum_{n=1}^{p}a^j_{0,n} \cos(n \phi)$, $j = 1,2$, where
$p$ can be a small integer, and $a^j_{0,n}=0$ for $n>p$ for
computational easiness. Note that our approach encompasses the case
of toroidal cloaks \cite{torus} but cloaks with irregular boundaries
such as cubes, fall beyond the scope of this study. Nevertheless,
contrarily to previous three dimensional numerical studies
\cite{torus,fengprl}, our scheme can be used in the design of cloaks
with non-convex boundaries.

To illustrate our methodology, we compute the electromagnetic field
diffracted by a cloak with rotational symmetry about the $z$-axis.
We perform full-wave finite element simulations in the commercial
package COMSOL,
when the cloak is illuminated by an approximate plane wave
(generated by a constant electric field on the upper surface of the
computational domain).
\section{Change of coordinates and pullbacks from optical to physical space}

\subsection{Material properties of the heterogeneous anisotropic cloak}
The  geometric transformation which maps the field within the full
domain $\rho\leq R_2(\phi)$ onto the annular domain $R_1(\phi)\leq
\rho'\leq R_2(\phi)$ can be expressed as:
\begin{equation}
\hspace{-0.1cm}\begin{cases} \rho'(\rho,\phi) =
\displaystyle{R_1(\phi)+\rho\frac{R_2(\phi)-R_1(\phi)}{R_2(\phi)}}
\; ,
 \cr
 \theta' = \theta \; , \; 0<\theta\leq 2\pi \; , \cr
 \phi' = \phi \; , \; -\pi/2<\phi\leq \pi/2
\end{cases}
\label{transfo_coordpendry}
\end{equation}
where $0\leq \rho\leq R_2(\phi)$. Note that the transformation maps
the field for $\rho > R_2(\phi)$ onto itself through the identity
transformation.

This change of co-ordinates is characterized by the transformation
of the differentials through the Jacobian:
\begin{equation}
\mathbf{J}(\rho',\phi')= \displaystyle{ \large
\frac{\partial(\rho(\rho',\phi'),\theta,\phi)}{\partial(\rho',\theta',\phi')}}
\; .
\end{equation}

This change of coordinates amounts to replacing a homogeneous
isotropic medium with scalar permittivity and permeability
$\varepsilon$ and $\mu$, by a metamaterial described by anisotropic
heterogeneous matrices of permittivity and permeability given by
\cite{opl2007,nicolet}
\begin{equation}
\underline{\underline{\varepsilon'}} =\varepsilon \mathbf{T}^{-1} \;
,  \quad
 \hbox{and} \quad
\underline{\underline{\mu'}}=\mu \mathbf{T}^{-1} \; , \label{epsmuT}
\end{equation}
where $\mathbf{T} \!= \! \mathbf{J}^T \mathbf{J}/\det(\mathbf{J})$
is a representation of the metric tensor in the so called stretched
radial coordinates. Note that there is no change in the impedance of
the media since the permittivity and permeability undergo the same
transformation.

After some elementary algebra, we find that

\begin{equation}
\mathbf{T}^{-1} =
\left( \begin{array}{ccc}
                \frac{c_{13}^2+\rho(\rho', \phi)^2}{c_{11}\rho'^2} & 0 & -\dfrac{c_{13}}{\rho'}\\
             0 & c_{11} & 0\\
             -\dfrac{c_{13}}{\rho'} & 0 & c_{11}\\
               \end{array}\right)
               \; ,
\label{invt}
\end{equation}
where
\begin{equation}
c_{11}(\phi') = \frac{R_2(\phi')}{R_2(\phi') - R_1(\phi')}
\; ,
\label{c11}
\end{equation}
and
\begin{equation}
c_{13}(\phi') = R_2(\phi')
\dfrac{\rho'-R_2(\phi')}{(R_2(\phi')-R_1(\phi'))^2} \dfrac{d
R_1(\phi')}{d\phi'} + R_1(\phi')
\dfrac{R_1(\phi')-\rho'}{(R_2(\phi')-R_1(\phi'))^2} \dfrac{d
R_2(\phi')}{d\phi'} \; ,
\label{c13}
\end{equation}
for $R_1(\phi')\leq\rho'\leq R_2(\phi')$. Elsewhere,
$\mathbf{T}^{-1}$ reduces to the identity matrix ($c_{11} =1$, $c_{13} =0$ and $\rho=\rho'$ for $\rho'>R_2(\phi')$).

\subsection{Singularity analysis of the transformation matrix}
\begin{figure}[h!]
\vspace{1.4cm}\mbox{}
\centerline{%
{\includegraphics[width=7cm]{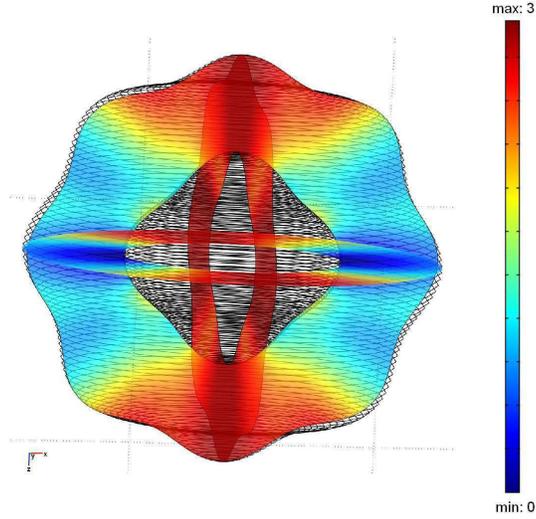}}} %
\caption{3D plot of ${(T^{-1})}_{11}$, as given by Eq. \ref{invt}, Eq. \ref{c11} and Eq. \ref{c13}
within the cloak with boundaries given by Eq. \ref{ddcloak}. The symmetries are noted.}
\label{t11}
\end{figure}

To exemplify the symmetric nature of the coefficients of $\mathbf{T}^{-1}$, and due
to the role played by its first entry in the following singularity
analysis, we show the variation of ${(T^{-1})}_{11}$ in Fig.
\ref{t11}. We note that this coefficient varies between zero and
three. The fine mesh on the inner boundary of the cloak is also
apparent from Fig. \ref{t11}, as are the non-convex and symmetric
features of the cloak.

In the case of a spherical cloak, $c_{13}$ vanishes and
$\mathbf{T}^{-1}$ reduces to
$\hbox{Diag}(\frac{\rho^2}{c_{11}\rho'^2},c_{11},c_{11})$. We note
that the first eigenvalue vanishes on the inner boundary and the
other two remain constant, which is consistent with the singularity
analysis led in \cite{kohn}. This is unlike the circular cylindrical
case whereby one eigenvalue goes to zero while the other one goes to
infinity on the inner boundary \cite{opl2007}.

However, in our case, the cloak is of an arbitrary shape, and it is
therefore illuminating to look at the behaviour of its permittivity
and permeability tensors's eigenvalues. The eigenvalues of
(\ref{invt}) are found to be
\begin{equation}
\begin{array}{ll}
\lambda_j=\frac{c_{13}^2+\rho^2+c_{11}^2\rho'^2}{2c_{11}\rho'^2}
+\frac{{(-1)}^j}{2}\sqrt{{\left(\frac{c_{13}^2+\rho^2+c_{11}^2\rho'^2}{c_{11}\rho'^2}\right)}^2
-4\frac{\rho^2}{\rho'^2}} \; , \; j=1,2 \; , \hbox{ and } \lambda_3=c_{11} \; .
\end{array}
\label{eiginvt}
\end{equation}
We can therefore see that $\lambda_j$, $j=1,2$ are spatially varying
functions of $\rho'(\rho,\phi)$, such that $\lambda_1=0$ when
$\rho=0$ i.e. at the inner boundary, while $\lambda_2>0$.
Importantly, $\lambda_3$ is a strictly positive constant for a given
angle $\phi'$ (i.e. a function independent upon $\rho$). Last, we
checked that when there is no longer a symmetry of revolution about
one axis, all three eigenvalues are also spatially varying with
$\theta'$, but $\lambda_3$ remains independent upon $\rho$.
This reflects the fact that the geodesics for light within the
arbitrary cloak follow more and more complex trajectories when we
perturb the geometry away from the spherical cloak.

\section{Finite elements computations}

\begin{figure}[h!]\centerline{%
{\includegraphics[width=6.cm]{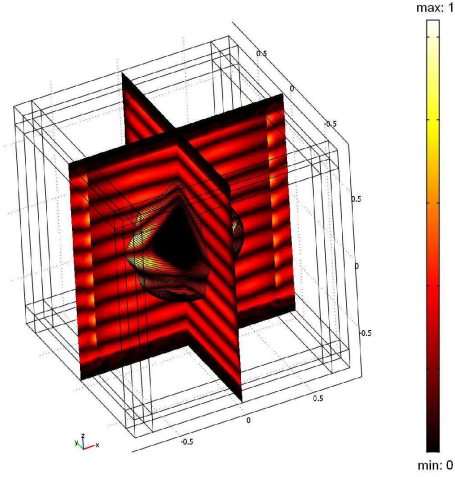}}} %
\caption{3D plot of the magnitude $\sqrt{E_1^2+E_2^2+E_3^2}$ of the
total electric field for a plane wave of wavenumber $k=2\pi/0.3$
incident from above on an non-convex invisibility cloak.}
\label{fig3d}
\end{figure}

We would now like to further investigate the electromagnetic
response of the cloak to an incident plane wave from above, see Fig.
\ref{fig3d}.
For this, we choose  the electric field ${\bf E}$ as the unknown:
\begin{equation}\label{eq:El}
\nabla\times\left( \underline{\underline{\mu'}}^{-1}\nabla
\times{\bf E} \right) - k^2\underline{\underline{\varepsilon'}} {\bf
E}={\bf 0}
\end{equation}
where $k=\omega\sqrt{\mu_0\varepsilon_0}=\omega/c$ is the
wavenumber, $c$ being the speed of light in vacuum, and
$\underline{\underline{\varepsilon'}}$ and
$\underline{\underline{\mu'}}$ are defined by Eqs. (\ref{epsmuT}).
Also, ${\bf E}={\bf E}_i+{\bf E}_d$, where ${\bf E}_i$ is the
incident field and ${\bf E}_d$ is the diffracted
 field which satisfies the usual outgoing wave conditions
(to ensure existence and uniqueness of the solution). The weak
formulation associated with Eq. \ref{eq:El} is discretised using
second order finite edge elements (or Whitney forms) which behave
nicely under geometric transforms (pull-back properties)
\cite{nicolet}.

For the sake of illustration, {\it cf.} Fig. \ref{fig3d}, let us
consider a cloak with inner and outer boundaries expressed as
\begin{equation}
R_1(\phi) =0.2+0.02\cos(4\phi) \; , \;
R_2(\phi) =0.4+0.02\cos(8\phi) \; . \;
\label{ddcloak}
\end{equation}

\begin{figure}[h!]\centerline{%
{\includegraphics[width=6cm,angle=0]{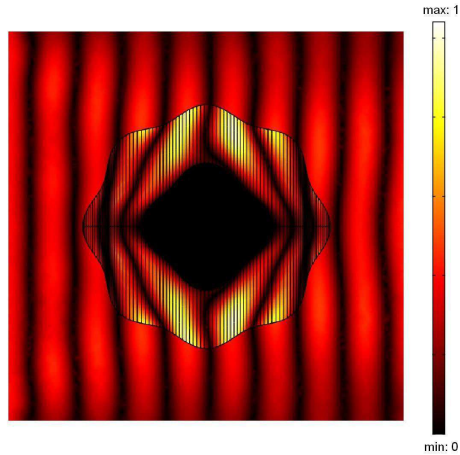}}} %
\vspace{0.0cm}\mbox{} \caption{2D plot of $\sqrt{E_1^2+E_2^2+E_3^2}$
generated by a slice of Fig. \ref{fig3d} in the $xz$-plane for
$y=0$.} \label{figexxz}
\end{figure}

\begin{figure}[h!]\centerline{%
{\includegraphics[width=7cm]{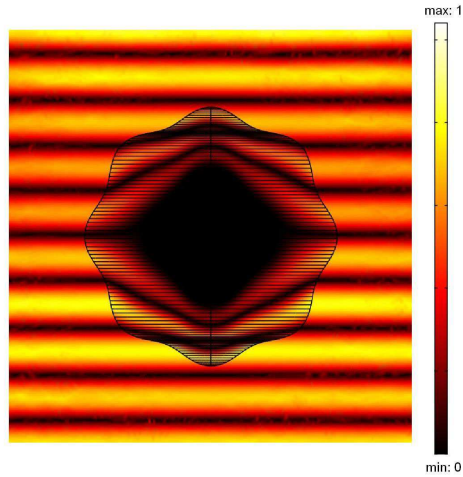}}} %
\vspace{0.0cm}\mbox{} \caption{2D plot of $\sqrt{E_1^2+E_2^2+E_3^2}$
generated by a slice of Fig. \ref{fig3d} in the $yz$-plane for
$x=0$.} \label{figexzy}
\end{figure}

We report the computations for the magnitude of the total electric
field in Fig. \ref{fig3d} (3D plot), Fig. \ref{figexxz} (2D plot in the
$xz$-plane) and Fig. \ref{figexzy} (2D plot in the $yz$-plane) for a
plane wave incident from above at wavenumber $k=2\pi/0.3$ (units are
in inverse of a length, say $\mu\hbox{m}^{-1}$ for nearly visible
light (UV)).

Around $4.10^5$ elements were used in this computation,
which corresponds to about $2.7 \; 10^6$ degrees of freedom. While
the convergence of the numerical scheme has been checked by
considering different types of meshes, the large size of the system
means we were not able to further refine the mesh for the
computational resources at hand.

\section{Conclusion}
In conclusion, we have proposed a design of an arbitrarily shaped
cloak using a Fourier approach. We only assumed that the cloak
displays a symmetry of revolution about one axis in order to reduce
the computational complexity of the problem (extension to completely
arbitrarily shaped cloaks is a straightforward matter).
Cloaking has been confirmed numerically for an incident plane wave
in resonance with the concealed region and an analysis of the
cloak's singularity has been carried out.

\noindent A. Diatta and S. Guenneau acknowledge funding from EPSRC grant
EP/F027125/1.

\end{document}